\documentclass[a4paper,10pt]{article}
\usepackage{amsfonts}
\usepackage{amsmath,amssymb}
\usepackage{latexsym}
\usepackage[english]{babel}
\usepackage{tikz}
\usepackage{amsthm}
\usepackage{hyperref}
\usepackage{natbib}

\makeatletter
\g@addto@macro\bfseries{\boldmath}
\makeatother

\numberwithin{equation}{section}

\begin{document}

\title{Non zero Coriolis field in Ehlers' Frame Theory}

\author{Federico Re$^1$, Oliver F. Piattella$^{2,3,4,5}$}

\maketitle

$^{1}$ \quad Dipartmento di Fisica Giuseppe Occhialini, Universit\`a di Milano-Bicocca, Piazza della Scienza 3, 20126, Milano, Italy; federico.re@unimib.it\\
$^{2}$ \quad Dipartimento di Scienza e Alta Tecnologia, Universit\`a degli Studi dell'Insubria, via Valleggio 11, I-22100 Como, Italy; of.piattella@uninsubria.it\\
$^{3}$ \quad INFN Sez. di Milano, Via Celoria 16, 20126, Milano, Italy.\\
$^{4}$ \quad N\'ucleo Cosmo-ufes, Universidade Federal do Esp\'irito Santo, avenida F. Ferrari 514, 29075-910 Vit\'oria, Esp\'irito Santo, Brazil.\\
$^{5}$ \quad Como Lake centre for AstroPhysics (CLAP), DiSAT, Università dell’Insubria, via Valleggio 11, 22100 Como, Italy.

\medskip
\textbf{Abstract.}\\
Ehlers' Frame Theory is a class of geometric theories parameterized by $\lambda := 1/c^2$ and identical to the General Theory of Relativity for $\lambda \neq 0$. The limit $\lambda \to 0$ does not recover Newtonian gravity, as one might expect, but yields the so-called Newton-Cartan theory of gravity, which is characterized by a second gravitational field $\boldsymbol{\omega}$, called the Coriolis field. Such a field encodes at a non-relativistic level the dragging feature of general spacetimes, as we show explicitly for the case of the $(\eta,H)$ geometries. Taking advantage of the Coriolis field, we apply Ehlers' theory to an axially symmetric distribution of matter, mimicking, for example, a disc galaxy, and show how its dynamics might reproduce a flattish rotation curve. In the same setting, we further exploit the formal simplicity of Ehlers' formalism in addressing non-stationary cases, which are remarkably difficult to be treated in the General Theory of Relativity. We show that the time derivative of the Coriolis field gives rise to a tangential acceleration which allows to study a possible formation in time of the rotation curve's flattish feature.


\section{Introduction}

Newtonian gravity constitutes an emblematic example in the history and philosophy of science of a theory that has been extended but not superseded. Indeed, Newtonian gravity is reliable within some regimes of applications, while out of these one employs the General Theory of Relativity (GR). Nevertheless, it is a noteworthy mathematical issue \citep{Einstein:1916vd,friedrichs1928invariante,dautcourt1964newtonske,hoffmann1966perspectives,kunzle1976covariant,ehlers1980isolated,Ehlers:2019aco} to precisely define in which sense Newtonian gravity is a limit of GR. In recent years, this theoretical issue has gained practical implications.

For the study of many astrophysical systems, GR is applied through the post-Newtonian (PN) expansion. However, in general, this paradigm is not mathematically equivalent to GR. The PN approach allowed us to achieve undoubtedly extraordinary successes, such as the explanation of the anomalous precession of the perihelion of Mercury \citep{Park_2017}, and the description of the coalescence of black holes, recently confirmed by the detection of their gravitational wave emission \citep{Abbott_2016,Antelis_2018}. On the other hand, many mathematical consequences of GR were discovered that cannot be described as PN corrections, thus being non-Newtonian phenomena related to the geometric nature and the non-linearity of GR. For example, geons \citep{anderson1997gravitational} and Carlotto-Shoen gravitational shields \citep{carlotto2016localizing}.

These are examples of gravitational systems that continue to exhibit non-Newtonian features even for weak gravitational forces and when any matter content has sub-relativistic speeds. Although they are just theoretical examples, still they prove how GR may not reduce to Newtonian gravity, even in a regime of low energy and low speeds. In other words, it is possible that not all the consequences of GR can be described by some term in a PN expansion. 

This fact of principle was pushed forward with the development of GR models of real astrophysical objects, wondering whether the PN description employed so far has missed some non-negligible aspect of the dynamics. This was made for the first time for the case of disc galaxies in \cite{Balasin:2006cg} and applied in the study of the Milky Way, finding a good fit with the observed rotation curve \citep{crosta2020, Crosta2023}. As recently shown in \cite{Seifert:2024wau}, and some time ago in \cite{Rizzi:2010yj}, certain vacuum solutions of GR can explain the flattish velocity profiles, due to their symmetry. The toy model of \cite{Balasin:2006cg} was improved by removing the unrealistic assumption of rigid rotation by \cite{Astesiano:2021ren, Re2023, Re:2024qco,GW}, formulating the so-called $(\eta, H)$ model. Further improvement was made in \cite{GRW}, considering a non-zero pressure source. In all these solutions the rotation is sustained not only by the centripetal attraction of the matter source, but also by a frame dragging effect, i.e. a non-negligible off-diagonal term in the metric matrix; the essentially non-Newtonian features of GR are thus exploited.

Criticism against this approach was raised; see, e.g., \cite{Ciotti:2022inn, Costa:2023awm, Glampedakis:2023qre, lasenby2023gravitomagnetism, barker2023does}. One of the main objections is that galaxy rotation attains speeds typically of order $10^{-3}\,c$, hence sub-relativistic, and the density of matter is far too low to produce strong gravitational potentials. For these reasons, the role of GR may be expected to be totally negligible. This claim is sustained by calculations performed with the linearized version of Einstein's equations, which are usually referred to as Gravitomagnetism and which can be framed in a PN expansion. The gravitomagnetic formalism is also employed by \cite{Ludwig:2021kea, AstesianoRuggiero1, AstesianoRuggiero2, Srivastava2023, LeCorre:2024tga}, with the crucial difference that these authors do not only evaluate the gravitomagnetic field generated by rotating matter, but also add the homogeneous solution of the field equations. Such a homogeneous term does not generally fit a PN expansion and is not subdominant with respect to the Newtonian gravitational potential.

The main issue is whether the linearized gravitomagnetic approach is suitable to describe a gravitational system such as a disc galaxy or whether some fundamentally non-Newtonian GR features can be involved. In other words: does GR really reduce to Newtonian gravity for such systems, characterized by weak fields and slow speeds? In this context, it is of extreme interest to consider \cite{Ehlers:2019aco}, the translated version of a paper by J. Ehlers in which Newtonian gravity is examined as the limit of GR.

Ehlers discovers a class of geometric theories, collectively named \textit{Frame Theory} (FT), indexed by a parameter $\lambda = 1/c^2$. For $\lambda \neq 0$, GR is recovered. Varying $\lambda$ has the physical meaning of considering dynamical systems with typical speeds closer to or much smaller than the speed of light. The limit $\lambda\rightarrow\infty$ thus returns the Carroll algebra, whereas the subrelativistic speed regime is mathematically obtained for the limit $c\rightarrow\infty$, equivalent to the choice $\lambda = 0$. Since any PN correction to Newton's equations is proportional to some power of $\lambda$, one expects to recover Newtonian theory for $\lambda = 0$. To Ehlers' own surprise, this is not the case.

The case $\lambda = 0$ is the so-called Newton-Cartan Theory (NC) \cite{ehlers1997examples,trautman2006einstein,NC_Costa,NC_Hartong}, which contains Newton gravity as a particular case. In fact, in addition to the usual Newtonian gravitational force field $\mathbf{g}$ (we adopt Ehlers' notation, boldface letters represent usual 3-vectors), or acceleration field, there is a second field $\boldsymbol{\omega}$, named after the way it enters the equation of motion as angular velocity field, or Coriolis field. Newton's theory is recovered only when $\boldsymbol{\omega} = 0$. 

Ehlers proved theorems (see Section 2 of \cite{Ehlers:2019aco}), where he shows how the Coriolis field vanishes under some suitable boundary conditions on hyperplanes of constant time. This asymptotic flatness condition represents an isolated gravitational system, to which Ehlers philosophically refers as ``real physical structures'' and ``measurable theories''. It was noted in \cite{BuchertFT} that the Coriolis field vanishes also under periodic boundary conditions, which are posed, e.g., for the spatially compact cosmologies developed in \cite{BuchertEhlers}.
However, $\boldsymbol{\omega}$ is, in general, present. It can be interpreted as what remains of the off-diagonal components of the metric when the $\lambda\rightarrow0$ limit is taken. Hence, it is strongly correlated with the frame dragging effect studied in \cite{Balasin:2006cg, crosta2020, Crosta2023, Astesiano:2021ren, Re2023, Re:2024qco,GW,GRW} and with the homogeneous gravitomagnetic term highlighted in \cite{Ludwig:2021kea, AstesianoRuggiero1, AstesianoRuggiero2, Srivastava2023, LeCorre:2024tga}.

This paper aims to push forward Ehlers' formalism, making explicit how the Coriolis field can constitute one of the GR phenomena that does not necessarily become negligible even in a low-energy regime. In Section \ref{Sec:Frametheory}, we present the equations of NC. In Section \ref{s3}, they are solved for the usual stationary, axisymmetric case. The $(\eta, H)$ non-rigid model for a disc galaxy is interpreted in terms of FT in Section \ref{s4}. The field equations of NC are much simpler than the full Einstein equations, so it is viable to study them even dropping some of the usual simplifying assumptions; stationarity is thus no longer assumed in Section \ref{s5}. In the concluding section, the advantages and disadvantages of FT are discussed.

\section{The Frame Theory and its $c\to\infty$ limit}\label{Sec:Frametheory}

The key point of Ehlers' FT is the existence of two metrics, the \textit{temporal} one $g_{\mu\nu}$ and the \textit{spatial} one $h^{\mu\nu}$. These are related by:\footnote{We use here Greek indices $\mu,\nu\dots$ to denote the coordinates from $0$ to $3$, while Latin indices $i,j,\dots$ will be used for the spatial coordinates $1,2,3$ alone.}
\begin{equation} \label{g h}
    g_{\mu\rho}h^{\rho\nu} = -\lambda \delta_{\mu}{}^\nu\,,
\end{equation}
where $\lambda := 1/c^2$. For $\lambda \neq 0$, the above relation is, a part from a minus sign, the standard one, establishing that $h$ is the inverse of $g$. However, when $\lambda = 0$ (that is, the non-relativistic limit $c \to \infty$), the two metrics have to be considered as independent tensors. In this case, Galilean local coordinates can be chosen such that:
\begin{equation}\label{Galcoords}
    g_{\mu\nu} = \delta^0{}_{\mu}\delta^0{}_{\nu}\,, \qquad h^{\mu\nu} = \delta^{\mu}{}_i\delta^{\nu}{}_j\delta^{ij}\,.
\end{equation}
Both metrics satisfy the compatibility, or metricity, condition (Ehlers' axiom 4 in \cite{Ehlers:2019aco}):
\begin{equation}\label{metrcond}
    g_{\mu\nu;\rho} = 0\,, \qquad h^{\mu\nu}{}_{;\rho} = 0\,,
\end{equation}
where the semicolon denotes covariant derivation with respect to a symmetric linear connection $\Gamma$ (a comma will denote usual partial derivative). These conditions, plus the choice of Galilean coordinates in Eq. \eqref{Galcoords}, provide constraints on the connection when $\lambda = 0$. In particular, the only non-zero components of $\Gamma$ are:
\begin{equation}\label{gOmegadef}
    \Gamma^i_{00} := -g^i\,, \qquad \Gamma^i_{j0} := \Omega_{ji} = -\Omega_{ij} = \varepsilon_{ijk}\omega^k\,.
\end{equation}
In the limit $\lambda \to 0$, and in Galilean coordinates, the geodesic equation reduces to:\footnote{If angular coordinates are used, $\Gamma^i_{jk}$ is not vanishing, in general.}
\begin{equation}
    \ddot x^i = -\Gamma^i_{00} - 2\Gamma^i_{0j}\dot x^j = g^i + 2\Omega_{ij}\dot x^j = g^i + 2\varepsilon_{ijk}\omega^k\dot x^j\,,
\end{equation}
or, in vector notation:
\begin{equation}\label{eqomotion}
    \ddot{\mathbf x} = \mathbf{g} + 2\dot{\mathbf{x}}\times\boldsymbol{\omega}\,.
\end{equation}
The field equations of NC can be found by taking the limit $\lambda \to 0$ of Einstein's equations and of the symmetry relations of the Riemann tensor. They are the following:\footnote{We find a contribution $2\omega^2$ instead of $\omega^2$ in the equation for $\nabla\cdot\mathbf{g}$. Probably, there is a typo in \cite{Ehlers:2019aco}.}
\begin{align}
   \label{omegaeqs} \nabla\times\boldsymbol{\omega} = 0\,, &\qquad \nabla\cdot\boldsymbol{\omega} = 0\,,\\
    \nabla\times\mathbf{g} + 2\dot{\boldsymbol{\omega}} = 0\,, &\qquad \nabla\cdot\mathbf{g} = -4\pi G\varrho + 2\omega^2\,. \label{geqs}
\end{align}
The term $2\dot{\mathbf{x}}\times\boldsymbol{\omega}$ is the well-known Coriolis acceleration, which appears in Newton's theory when writing the acceleration of a point particle in a rotating reference frame. For this reason $\boldsymbol{\omega}$ is dubbed the\textit{ Coriolis field} by Ehlers in \cite{Ehlers:2019aco}. However, differently from the fictitious force in Newton's dynamics, $\boldsymbol{\omega}$ is here a physical field to be treated on the same footing as $\mathbf{g}$. In particular, $\boldsymbol{\omega}$ is the antisymmetric part of the gravitational field that is left behind from the limit $\lambda \to 0$ and that is not necessarily subdominant with respect to $\mathbf{g}$, as is the case in a standard post-Newtonian expansion \cite{Weinberg:1972kfs}.

Note that the Coriolis field is not sourced from the matter content. However, it can modify the local dynamics depending on the boundary conditions to which it is subjected. Moreover, note also that, by Helmholtz's theorem, if $\boldsymbol{\omega}$ is analytic and vanishes at infinity, then $\boldsymbol{\omega} = 0$. So, if we want to discuss a non-trivial $\boldsymbol{\omega}$ we need to admit that it might be non-zero at infinity.

Since $\boldsymbol{\omega}$ is both solenoidal and irrotational, it can be expressed as the gradient of a harmonic scalar function $V$, that we dub \textit{Coriolis potential}, that is:
\begin{equation}
    \boldsymbol{\omega} = \nabla V\,, \qquad \nabla^2V = 0\,.
\end{equation}
Note that if $\boldsymbol{\omega}$ is spatially homogeneous, it can be eliminated via a coordinate transformation to a rotating frame, and the theory becomes identical to Newton's. Indeed, write the equation of motion as:
\begin{equation}
    \ddot{\mathbf x} = \mathbf{g}' - 2\boldsymbol{\omega}\times\dot{\mathbf{x}} - \boldsymbol{\omega}\times(\boldsymbol{\omega}\times\mathbf{x}) - \dot{\boldsymbol{\omega}}\times\mathbf{x}\,,
\end{equation}
where:
\begin{equation} \label{g'}
    \mathbf{g}' := \mathbf{g} + \boldsymbol{\omega}\times(\boldsymbol{\omega}\times\mathbf{x}) + \dot{\boldsymbol{\omega}}\times\mathbf{x}\,.
\end{equation}
So we have:
\begin{equation}
    \ddot{\mathbf x}' := \ddot{\mathbf{x}} + 2\boldsymbol{\omega}\times\dot{\mathbf{x}} + \boldsymbol{\omega}\times(\boldsymbol{\omega}\times\mathbf{x}) + \dot{\boldsymbol{\omega}}\times\mathbf{x} = \mathbf{g}'\,,
\end{equation}
with $\ddot{\mathbf x}'$ the acceleration experienced in a rotating frame. Now, noting that for a \textit{homogeneous} $\boldsymbol{\omega}$ one has:
\begin{align}
    &\nabla \times (\dot{\boldsymbol{\omega}} \times \mathbf{x}) = 2\dot{\boldsymbol{\omega}}\,, \qquad \nabla \cdot (\dot{\boldsymbol{\omega}} \times \mathbf{x}) = 0\,,\\
    &\nabla \times [\boldsymbol{\omega} \times (\boldsymbol{\omega} \times \mathbf{x})] = 0\,, \qquad \nabla \cdot [\boldsymbol{\omega} \times (\boldsymbol{\omega} \times \mathbf{x})] = -2\omega^2\,, 
\end{align}
the field equations become:
\begin{align}
    \nabla\times\mathbf{g}' = 0\,, \quad \nabla\cdot\mathbf{g}' = -4\pi G\varrho\,,
\end{align}
which, together with $\ddot{\mathbf x}' = \mathbf{g}'$ are Newton's gravity.

On the other hand, an inhomogeneous $\boldsymbol{\omega}$ cannot be absorbed in this way and therefore must be considered as an independent field.

NC descends from GR, thus implying local freedom of movement for the reference frames. The physical Coriolis field emerges from this freedom: it describes a rotation of the reference frame, varying, in general, from point to point. For this reason, it cannot be absorbed with a global gauge. The particular case of a spatially homogeneous Coriolis field describes an observer that is rotating with the same angular speed everywhere, i.e. is rigidly rotating, and hence its effects on the dynamics can be trivially interpreted by recognizing that a rotating frame was chosen.\\
Each of the $\boldsymbol{\omega}$ terms in Eqs. (\ref{eqomotion}, \ref{geqs}) can indeed be interpreted as emerging from an apparent acceleration in Galilean relativity, having nevertheless a spatially inhomogeneous $\boldsymbol{\omega}(\textbf{x})$. The $2\omega^2$ additional term in the modified Poisson equation is a manifestation of ``local'' centrifugal acceleration $\textbf{a}_c(\textbf{x}):=\boldsymbol{\omega}(\textbf{x})\times(\boldsymbol{\omega}(\textbf{x})\times\textbf{x})$, which has no consequences for $\nabla\times\textbf{g}$ since it is irrotational in $\textbf{x}$. On the other hand, the azimuthal acceleration $\textbf{a}_a(\textbf{x}):=\dot{\boldsymbol{\omega}}(\textbf{x})\times\textbf{x}$ is divergenceless in $\textbf{x}$, returning instead a source $2\dot{\boldsymbol{\omega}}$ for the curl of $\textbf{g}$. Inertial acceleration $\textbf{a}_i(\textbf{x})$ does not play a role here, since it comes from the translational motion of the coordinate frame instead of the rotational one, thus having no dependence on $\boldsymbol{\omega}$. The last apparent acceleration, according to the Galilean transformation law:
\begin{equation}
    \textbf{g}' = \textbf{g}(\textbf{x}) + \textbf{a}_i(\textbf{x}) + \textbf{a}_c(\textbf{x}) + \textbf{a}_C(\textbf{x}, \dot{\textbf{x}}) + \textbf{a}_a(\textbf{x})\,,
\end{equation}%
is the Coriolis acceleration $\textbf{a}_C = 2\boldsymbol{\omega}(\textbf{x})\times\dot{\textbf{x}}$. Unlike the others, it cannot be absorbed by redefining the gravitational acceleration $\textbf{g}(\textbf{x})$, since the latter must depend only on $\textbf{x}$, and not also on the velocity field $\dot{\textbf{x}}$. For this reason, it does not appear in the transformation (\ref{g'}) and its consequence is a correction term in the Equations of Motion (\ref{eqomotion}), instead of in the Field Equations (\ref{geqs}).

We note that it is not prohibited to redefine a more general, velocity-dependent acceleration field $\textbf{g}'(\textbf{x},\dot{\textbf{x}})$. However, such a $\textbf{g}'(\textbf{x},\dot{\textbf{x}})$ would not be interpretable as a Newtonian gravitational field, since in the Newtonian paradigm it depends only on the position. This $\textbf{g}'(\textbf{x},\dot{\textbf{x}})$ would be instead a gravitational analogue of the electromagnetic Lorentz force. The velocity dependence of magnetic force opened, indeed, the problem of which is the correct classical reference frame where to measure this velocity, leading eventually to the birth of Special Relativity. Incorporating the Coriolis acceleration analogue in the field $\textbf{g}'$ would pose a similar problem. Indeed, as remarked earlier, the very existence of the Coriolis field descends from the freedom of choosing any reference frame in GR, which is partially inherited by the non-relativistic Newton-Cartan theory of gravity.

Because of this analogy with apparent accelerations in Galilean relativity, one could wonder if even an inhomogeneous Coriolis field may be just a gauge artifact. If we see NC as the low-energy regime of GR, the arbitrariness in the choice of coordinates also allows for a local gauge freedom. Can this be used to absorb the Coriolis field in general? This is not the case, as can be verified by calculating suitable scalar quantities that depend on the Coriolis field. For instance, we can write explicitly the Kretschmann scalar as%
\begin{equation}
    K = -8\lambda||\underline{\underline{J}}\boldsymbol{\omega}||^2\,,
\end{equation}%
where $\underline{\underline{J}}\boldsymbol{\omega}$ denotes the Jacobian matrix of $\boldsymbol{\omega}(\textbf{x})$, of which the norm is taken in the matrix space. This can be immediately proved from Eqs. (16) and (17) of \cite{Ehlers:2019aco}, by contracting the indices $d$ and $e$ through (\ref{g h}). It should be noted that the Kretschmann scalar depends on all and only the first derivatives of $\boldsymbol{\omega}$, so that any spatially homogeneous term of the Coriolis field does not affect it, being a coordinate artifact.

In the following, we refer to the case $\boldsymbol{\omega} = 0$ as the \textit{Newtonian case}. In this instance we recover the usual set of equations:
\begin{align}
    \ddot{\mathbf x} = \mathbf{g}\,, \quad \nabla\times\mathbf{g} = 0\,, \quad \nabla\cdot\mathbf{g} = -4\pi G\varrho\,.
\end{align}
The irrotationality of $\mathbf{g}$ allows us to write it as the gradient of a potential $U$, that is, $\mathbf{g} = - \nabla U$. This potential satisfies the Poisson equation $\nabla^2U = 4\pi G\varrho$, whose general solution assuming vanishing fields at infinity is:
\begin{align}
    U = -G\int d^3\mathbf{x}'\frac{\varrho(\mathbf x')}{|\mathbf{x}' - \mathbf{x}|}\,.
\end{align}
If $\boldsymbol{\omega}$ does not vanish and is time-independent, we still can write $\mathbf{g} = - \nabla U$, but the potential is now determined by a \textit{modified Poisson equation}, whose solution can be written as follows again assuming vanishing fields at infinity:
\begin{align} \label{U bound}
    U = -G\int d^3\mathbf{x}'\frac{\varrho(\mathbf x')}{|\mathbf{x}' - \mathbf{x}|} + \frac{1}{4\pi}\int d^3\mathbf{x}'\frac{2\omega^2(\mathbf x')}{|\mathbf{x}' - \mathbf{x}|}\,.
\end{align}
This means that the square modulus of the Coriolis field contributes effectively as a negative matter density; therefore, it might seem that its presence would make more matter necessary to justify a given equivalent Newtonian dynamics. However, $\boldsymbol{\omega}$ also enters the equation of motion \eqref{eqomotion} in a way that might correspond to the necessity of a smaller matter density. So, one has to see which of the two effects is dominant \citep{Re:2024qco}.
   
\section{The stationary case and axial symmetry} \label{s3}

In his work \cite{Ehlers:2019aco, ehlers1997examples}, J. Ehlers applied his formalism to some noteworthy examples: the Schwarzschild metric, as well as the generic spherically symmetric metric; the Kerr metric; the plane gravitational wave; the Friedmann-Lemaître cosmology; the Gödel universe; and the NUT spacetimes. With the only exception of the NUT spacetimes, in the above examples the off-diagonal components of the metric display vanishing effects in the $\lambda\rightarrow 0 $ limit, i.e., their low-energy regime coincides with the Newtonian regime. It is noteworthy that studies on the Taub-NUT spacetimes involved some of the key concepts arising in the discussion about dragging in galaxy models, e.g. its interpretation in terms of gravitomagnetism, of the Coriolis field, and/or in terms of frame dragging itself: see \cite[pages 219 - 221]{griffiths2009exact}.

We are interested in considering here physical systems that are not Newtonian for low energies. In other words, we want to study a realistic example of NC with a non-zero Coriolis field. As we mentioned in the Introduction, a discussion in the literature related to this issue is about the dynamics of disc galaxies \citep{Balasin:2006cg, Crosta2023, Astesiano:2021ren, Re2023, Re:2024qco, GW, GRW, Ciotti:2022inn, Costa:2023awm}. Since we have in mind such a framework to apply Ehlers' FT, let us employ cylindrical coordinates $(r,\phi,z)$ with origin in the galactic center.\\
In the case of time-independent (stationary) fields one has $\dot{\boldsymbol{\omega}} = 0$, so the curl of $\mathbf{g}$ is vanishing and this field can be written as the gradient of a scalar potential $U$: 
\begin{align}
    \mathbf{g} = g_r\hat r + g_\phi\hat\phi + g_z\hat z = -\nabla U = -U_{,r}\hat r - \frac{1}{r}U_{,\phi}\hat\phi - U_{,z}\hat z\,.
\end{align}
Similarly for the Coriolis field:
\begin{align}
    \boldsymbol{\omega} = \omega_r\hat r + \omega_\phi\hat\phi + \omega_z\hat z = \nabla V = V_{,r}\hat r + \frac{1}{r}V_{,\phi}\hat\phi + V_{,z}\hat z\,.
\end{align}
In addition to stationarity, we demand axial symmetry. Therefore, no component of the fields may depend on $\phi$ and, consequently, the potentials cannot depend on $\phi$. We then have:
\begin{equation}
    \omega_\phi = g_\phi = 0\,.
\end{equation}
The position vector and its first and second derivatives are:
\begin{align}
    &\mathbf{x} = r\hat r + z\hat z\,, \quad \dot{\mathbf{x}} = \dot r\hat r + r\dot\phi\hat\phi + \dot z\hat z\,, \quad \ddot{\mathbf{x}} = (\ddot r - r\dot\phi^2)\hat r + (r\ddot\phi + 2\dot r\dot\phi)\hat\phi + \ddot z\hat z\,. \label{x dot dot}
\end{align}
The cross product between the velocity and the Coriolis field is:
\begin{equation}
    \dot{\mathbf{x}}\times\boldsymbol{\omega} = (r\dot\phi\omega_z)\hat r + (\dot z\omega_r - \dot r\omega_z)\hat\phi + (-r\dot\phi\omega_r)\hat z\,.
\end{equation}
The equation of motion can be then written in components as:
\begin{align}
    \ddot r - r\dot\phi^2 &= g_r + 2r\dot\phi\omega_z\,, \label{EoM r}\\
    r\ddot\phi + 2\dot r\dot\phi &= 2(\dot z\omega_r - \dot r\omega_z)\,,\\
    \ddot{z} &= g_z - 2r\dot\phi\omega_r\,. \label{EoM z}
\end{align}
The equations for the potentials are, using the expression for the Laplacian in cylindrical coordinates:
\begin{align}
    \frac{1}{r}\left(rV_{,r}\right)_{,r} + V_{,zz} &= 0\,,\\
    \frac{1}{r}\left(rU_{,r}\right)_{,r} + U_{,zz} &= 4\pi G\varrho - 2\left(V_{,r}^2 + V_{,z}^2\right)\,. \label{Poiss axisym}
\end{align}
We now solve the Laplace equation for the Coriolis potential and obtain a solution for the rotational velocity in the special case of a circular motion on the galactic plane.

\subsection{Zero separation constant solution for the Coriolis potential}

Using variable separation, and assuming a vanishing separation constant, a solution of the Laplace equation can be found with the following form:
\begin{equation}
    V_{0} = a_0(1 + b_0 z)(1 + c_0\ln r)\,,
\end{equation}
so that the Coriolis field is:
\begin{align}
    \omega_r = \frac{a_0c_0(1 + b_0z)}{r}\,, \qquad \omega_z = a_0b_0(1 + c_0\ln r)\,.
\end{align}
In order to avoid divergence in $r = 0$ we need $c_0 = 0$, and the only possible solution is:
\begin{align}
    \omega_z = \omega = a_0b_0\,.
\end{align}
As discussed in the previous Section, a spatially homogeneous Coriolis field is not a physical field, since it can be eliminated with a change of coordinates. In the present case, one can recognize that any effect from this constant Coriolis field is an apparent one, due to the choice of a non-inertial frame of coordinates $\textbf{x}$.

This example is in part related to the solution found in \cite{LeCorre:2024tga}, where the author works within the gravitomagnetic framework in the weak-field limit (thus neglecting the additional $2\omega^2$ term in the Modified Poisson Equation). It is not a spatially homogeneous solution, being in fact equivalent to a Coriolis field $\boldsymbol{\omega}(\textbf{x}) \propto (k_0 + K_1/r^2)\hat{z}$. Thus, it is a physical field. However, the constant term $k_0$ can be simplified by a change in coordinates.

\subsection{General solution for the Coriolis potential}

Using again variable separation, a general solution for the Coriolis potential is:
\begin{equation} \label{V gamma}
    V_{\gamma} = [a_\gamma J_0(\gamma r) + b_\gamma Y_0(\gamma r)][c_\gamma  \cosh(\gamma z) + d_\gamma \sinh(\gamma z)]\,,
\end{equation}
where $\gamma^2 \neq 0$ is the separation constant and $J_0,Y_0$ are, respectively, the Bessel functions of first and second kind of zeroth order. We look for a Coriolis field that does not diverge at $r = 0$. Preferably, we would like it to be vanishing in the galactic bulge because we know that there Newton's gravity works fine. Therefore, we need $b_\gamma = 0$. The Coriolis potential then becomes:
\begin{equation}
    V_{\gamma} = J_0(\gamma r)[c_\gamma\cosh(\gamma z) + d_\gamma \sinh(\gamma z)]\,.
\end{equation}
From this potential we have a Coriolis field:
\begin{align}
    &\omega_r = \gamma J'_0(\gamma r)[c_\gamma\cosh(\gamma z) + d_\gamma \sinh(\gamma z)]\,, \quad \omega_z = \gamma J_0(\gamma r)[c_\gamma\sinh(\gamma z) + d_\gamma \cosh(\gamma z)]\,, \label{single harmonic}
\end{align}
where the prime denotes derivation with respect to the argument. Note that $J_0(\gamma r)$ is 1 with zero derivative in $r = 0$ and goes to zero for large $r$ as $1/r^{1/2}$, if $\gamma$ is real. However, unfortunately, the behavior along $z$ is exponentially divergent. 

We could do as in \cite{Balasin:2006cg} and let substitute an imaginary $\gamma\rightarrow\gamma i$ parameter, so that we have: 
\begin{align}
    &\omega_r = \gamma I'_0(\gamma r)[c_\gamma\cos(\gamma z) + d_\gamma \sin(\gamma z)]\,, \quad \omega_z = \gamma I_0(\gamma r)[-c_\gamma\sin(\gamma z) + d_\gamma \cos(\gamma z)]\,.
\end{align}
Here, $I_0(\gamma r)$ is 1 with zero derivative in $r = 0$, but diverges as $e^r/\sqrt{r}$ for large radii. In this case, the behavior for $r \to \infty$ is not realistic, besides the fact that the potential does not vanish for large $z$.

Since both the choices of a real or an imaginary $\gamma$ lead to exponential divergences at spatial infinity, a realistic profile for the Coriolis field around a physical galaxy should be a suitable superposition of the $V_{\gamma}$ harmonics in (\ref{V gamma}). A similar construction was performed by \cite{Balasin:2006cg}.\\
Moreover, the boundary conditions for the Coriolis field may allow non-trivial choices. As is customary in physics, one would demand the vanishing of the field at infinity. On the other hand, one must take into account the fact that Ehlers' theory in the limit $c \to \infty$ is necessary a local one: it cannot describe the entire universe. For this reason, we pay attention only to avoid divergences for $r = 0$, but accept possible divergent behaviors for $r,z\to\infty$. For the case of a galaxy, this means to recognize that the metric outside is not necessarily a Minkowskian one, since it is surrounded by other galaxies and, at the spatial infinity, it should rather be approximable to the Friedmann-Lema\^itre-Robertson-Walker geometry. Moreover, the intergalactic space may be described by non-trivial vacuum solutions, e.g. as investigated in \cite{Rizzi:2010yj, Seifert:2024wau}, rather than the Minkowski metric.\\
The crucial role of non-zero boundary conditions in returning a non-vanishing Coriolis field was already clear in the seminal paper \cite{Ehlers:2019aco}, from his Theorems 1 and 2. Ehlers relates them to the concepts of isolated systems, and thus ``real physical structures'', in an epistemological sense.

Given the nice behavior of the Coriolis potential as a function of $r$, let us concentrate on the solution with $\gamma^2 > 0$.

\subsection{Circular motion on the galactic plane} \label{s3c}

Using the simple single-harmonic solution (\ref{single harmonic}) for the Coriolis field, let us write down the equation of motion for a $r = $ constant solution constrained on the galactic plane ($z = 0$):
\begin{align}
    - r\dot\phi^2 = g_r|_{z = 0} + 2r\dot\phi\omega_z|_{z = 0}\,, \label{geo use}\\
    r\ddot\phi = 0\,,\\
    \ddot{z} = g_z|_{z = 0} - 2r\dot\phi\omega_r|_{z = 0}\,.
\end{align}
In this case $r\dot{\phi} := v$ is a constant rotational velocity.

In order to have no runaway of particles from the equatorial plane, we suppose $g_z|_{z = 0} = -U_{,z}|_{z = 0} = 0$ and $\omega_r|_{z = 0} = 0$, which requires $c_\gamma = 0$. The condition on $\mathbf{g}$ is meaningful once we ask for some planar symmetry with respect to the $z=0$ plane. However, the condition on $\boldsymbol{\omega}$ does not descend from the imposing of the same planar symmetry on it. Neither we are interested in considering a planar symmetric Coriolis field, otherwise we would not be able to recover the galactic models of the form (\ref{eta H}).

We want thus to impose the planar symmetry $\varrho(r,z)=\varrho(r,-z)$ on the matter distribution, first of all, since it can be observationally checked for a real galaxy. Looking at the modified Poisson equation in its axisymmetric form (\ref{Poiss axisym}), we can choose to realize the symmetry by imposing it on each term. This means that the Newtonian potential $U(r,z)$ must be analogously a symmetric function in $z$; so that by evaluating $g_z$ in $z=0$ one must find zero. Meanwhile, the Coriolis potential $V(r,z)$ is allowed to be a symmetric or an antisymmetric function (depending on $z$). Choosing an antisymmetric $V$, we find the above condition $\omega_r(r,0)=0$ and this forces $\omega_z$ to be antisymmetric, so it is allowed to have $\omega_z(r,0)\neq0$.

Since we are limiting ourself to a Coriolis field with the form (\ref{single harmonic}), on the symmetry plane it is:
\begin{equation}
    \boldsymbol{\omega}|_{z=0} = \omega_z(r,0)\hat{z} = \gamma d_{\gamma} J_0(\gamma r)\hat{z}\,.
\end{equation}
We are therefore left to determine the shape of $U(r,z)$ near $z=0$, using the remaining equations (\ref{geo use}, \ref{Poiss axisym}). We can call $U_{,r}(r,0) := a_c(r)$ the centripetal acceleration and $U_{,zz}(r,0) := b(r)$. From the first one we have:
\begin{equation}
    v(r)^2 = r a_c(r) - 2rv(r) \gamma d_{\gamma} J_0(\gamma r)\,,
\end{equation}%
where the rotation profile $v(r)$ can be observed, and the second equation is:
\begin{equation}
    \frac{1}{r}\left(ra_c(r)\right)_{,r} + b(r) = 4\pi G\varrho(r,0) - 2\gamma^2 d_{\gamma}^2 J_0(\gamma r)^2\,.
\end{equation}
For $r$ small, $\omega_z|_{z = 0} \sim \gamma d_\gamma$, whereas $b$ and $\varrho|_{z = 0}$ tend to constants. So, we expect to reproduce the Newtonian case $v \sim r$.

On the other hand, for sufficiently large $r$, far from the galactic bulge, the density is negligible and one can obtain a flat rotation curve $v(r)\sim v_f$ for:
\begin{align}
    a_c(r) \sim &2v_f \gamma d_\gamma J_0(\gamma r) + \frac{v_f^2}{r} \sim v_f d_{\gamma}\sqrt{\frac{8\gamma}{\pi r}}\cos\left(\gamma r - \frac{\pi}{4}\right)\,, \\
    b(r) \sim & -2v_f \gamma d_{\gamma} \frac{J_0(\gamma r) + \gamma r J'_0(\gamma r)}{r} - 2\gamma^2 d_{\gamma}^2 J_0(\gamma r)^2 \sim v_f d_{\gamma}\sqrt{\frac{8\gamma^3}{\pi r}}\sin\left(\gamma r -\frac{\pi}{4}\right)\,.
\end{align}
However, we must remember that the shape (\ref{single harmonic}) has no zero boundary conditions, thus not producing a Newtonian potential of the form (\ref{U bound}). The freedom we had here for $v(r), a_c(r)$ and $b(r)$ comes from not having imposed any boundary conditions at spatial infinity. For a realistic galaxy, which is not a totally isolated system, instead being surrounded by other similar galaxies, we can imagine that $\boldsymbol{\omega}$ should be imposed to be non-zero, but neither exponentially growing, on the boundary. As we mentioned above, this should be realized by suitably superimposing many single-harmonic solutions (\ref{single harmonic}).

\section{The low energy limit of the dragging metric} \label{s4}

In \cite{Astesiano:2021ren} the galactic dynamics is modeled using the following stationary, axially, and planar symmetric metric (the\\
Lewis–Papapetrou–Weyl metric):
\begin{align}
    g_{\mu\nu}dx^{\mu}dx^{\nu} = &+e^{2\lambda U}(dt + \lambda\L_D d\varphi)^2 - \lambda e^{-2\lambda U}\left[r^2 d\varphi^2 + e^{2\lambda k}(dr^2 + dz^2)\right]\,. \label{eta H}
\end{align}
Here $U(r, z)$ is the Newtonian potential, $\L_D(r, z)$ is the quasilocal angular momentum (per unit mass) of spacetime, and $k(r, z)$ is a conformal spatial factor. This metric is suitable not only for a pressureless source, as is assumed in \cite{Astesiano:2021ren,Re2023,Re:2024qco,GW}, but also for the non-zero pressure case considered in \cite{GRW}, once one generalizes the component $g_{rr}=-\lambda e^{-2\lambda U}W(r, z)^2$, where the field $W$ tends to $r$ for $\lambda\to0$. \\
In the aforementioned literature, the full Einstein equations are cast, and the exact $(\eta, H)$ solution is explicitly given in the pressureless case. The subrelativistic regime of a real galaxy is subsequently imposed on the full GR equations, expanding each one at the lowest order in $v/c$, where $v\sim10^{-3}c$ is the typical speed of the system. This operation was performed in \cite{Astesiano:2021ren,Re2023} and made explicit in \cite{Re:2024qco}. For such a low-energy regime, \cite{GRW} found Field Equations (17 - 20) and Equations of Motion (21 - 22) that are not Newtonian, displaying indeed additional terms in $\L_D$.\\
Since for the $c\rightarrow\infty$ limit the Newtonian theory is not recovered, one must conclude that the PN expansion fails in this case. In other words, the spacetime metric cannot be described as a Minkowskian metric with higher-order corrections. To say that the Minkowskian spacetime cannot be taken as background metric means that the suitable background is a non-trivial one, carrying on some intrinsic energy and angular momentum. For this reason, \cite{GRW} interpreted $\L_D$ as the quasilocal angular momentum of this background spacetime, and the additional term in the Poisson equation (17) as its quasilocal energy.

The laborious procedure of casting the full GR equations and then expanding them in the low-energy limit can be substituted by the more straightforward NC, whose equations have already incorporated the low-energy regime. For this reason, the $\lambda$ factor has been restored in (\ref{eta H}). Applying Ehlers' formalism and taking the limit $\lambda\rightarrow0$, we recognize that:
\begin{equation}
\label{fields in eta H}
    \textbf{g} = -U_{,r}\hat{r} - U_{,z}\hat{z}\,, \qquad \boldsymbol{\omega} = \frac{\L_{D,z}\hat{r} - \L_{D,r}\hat{z}}{2r}\,.
\end{equation}
For the field equations, the first of (\ref{geqs}) is identically zero, since we have assumed stationarity ($\dot{\boldsymbol{\omega}}= 0$) and $\mathbf{g}$ is a gradient.

The modified Poisson equation becomes:
\begin{equation}
\label{Poiss mod}
    4\pi G\varrho = 2\omega^2 - \nabla\cdot\textbf{g} = \nabla^2 U + \frac{|\nabla \L_D|^2}{2r^2}\,,
\end{equation}
which is identical to the modified Poisson equation (17) presented in \cite{GRW}, where it is derived as the low energy limit of the Einstein equations. Here, it is proved in a more straightforward way, thanks to Ehlers' FT.

Analogously, we can recognize the other two field equations (\ref{omegaeqs}) as the low energy limits of Eqs. (18-20) of \cite{GRW}. Indeed, from Eq. (\ref{fields in eta H}) we find:
\begin{align}
    \nabla\times\boldsymbol{\omega} &= (\omega_{r,z} - \omega_{z,r})\hat{\varphi} =\left(\frac{\L_{D,zz}}{2r} + \frac{1}{2}\frac{\L_{D,rr}r - \L_{D,r}}{r^2}\right)\hat{\varphi}
    = \frac{\hat{\Delta}\L_D}{2r}\hat{\varphi}\,,
\end{align}
where $\hat{\Delta}$ is the so-called Grad-Shafranov differential operator \cite{grad1958hydromagnetic, shafranov1966equilibrium}. Here we can appreciate how such an unusual operator as the Grad-Shafranov one emerges here because of the $r$ at the denominator and the switched partial derivatives in the $\omega$-$\L_D$ relation (\ref{fields in eta H}): equation (18) of \cite{GRW}, namely the Grad-Shafranov equation for $\L_D$, is equivalent to requiring that $\boldsymbol{\omega}$ is an irrotational field.

Finally, the relation $\nabla\cdot\boldsymbol{\omega}$ can be obtained from equations (19-20) of \cite{GRW} by demanding the integrability condition $\partial_z k_{,r} = \partial_r k_{,z}$. In fact we have from (\ref{fields in eta H}) that (19-20) of \cite{GRW} become:
\begin{align}
    &k_{,r} = \frac{\L_{D,z}^2 - \L_{D,r}^2}{4r} = r\omega_r^2 - r\omega_z^2\,, \quad k_{,z} = \frac{\L_{D,r}\L_{D,z}}{2r} = 2r\omega_r \omega_z\,,
\end{align}
and the integrability condition takes the form:
\begin{align}
    \nabla\cdot\boldsymbol{\omega} &=  \omega_{r,r} + \frac{\omega_r}{r} + \omega_{z,z} = \frac{\L_{D,zr}r - \L_{D,z}}{2r^2} + \frac{\L_{D,z}}{2r^2} - \frac{\L_{D,rz}}{2r} = 0\,.
\end{align}
In other words, the Coriolis field is automatically solenoidal once defined as in (\ref{fields in eta H}), and the integrability condition in $k$ implies just that $\omega_r(\nabla\times\boldsymbol{\omega})_{\varphi} = 0$, which is already guaranteed by the Grad-Shafranov equation in \L$_D$.

We have thus proved that Ehlers' field equations (\ref{omegaeqs}) and (\ref{geqs}) are implicitly contained in Einstein's field equations in the low-energy limit considered in \cite{GRW} and that the residue of the quasilocal angular momentum $\L_D$ in the low-energy limit is tightly related to a non-zero Coriolis field. The \L$_D$ profiles proposed in the plot of $v_D$ of \cite{GRW} to describe flattish rotation curves can now be interpreted as superpositions of single-harmonic solutions with the form (\ref{single harmonic}).
We can thus write explicitly the Coriolis potential as:
\begin{equation}
    V(r,z)=\frac{1}{2\pi}\int dk \hat{\omega}(k) J_0(kr)\sinh{(kz)}\,,
\end{equation}
where the $c_{\gamma}$ coefficients in (\ref{single harmonic}) vanish, in order to have $\omega_r = V_{,r} =|_{z=0} 0$, and the Bessel-Fourier coefficients $\hat{\omega}(k)$ are given by:
\begin{equation}
    \hat{\omega}(k) = \int r dr \omega_z(r,0) J_0(kr)\,.
\end{equation}
Finally, here we can substitute the $V_{,z}(r,0) = \omega_z(r,0)$ profile by following the correspondence rule (\ref{fields in eta H}) and equation (36) of \cite{GRW}:
\begin{equation}
    \omega_z(r,0) = -\frac{\L_{D,r}}{2r} = -\frac{1}{2r}[rv_{obs}(r)]_{,r} + \frac{1}{2}\sqrt{\frac{1}{r^2}[rv_{obs}(r)]_{,r}^2 - \frac{2}{r}[v_h(r)^2]_{,r}}\,.
\end{equation}
Here, $v_{obs}(r)$ is the observed rotation curve for stars and gas on the galactic plane, and $v_h(r)$ is the halo correction of the rotation curve which is usually assumed in the $\Lambda$CDM paradigm. Such a Coriolis field $\boldsymbol{\omega}(r,z)$, if present, would thus explain the flattish profile of $v_{obs}$ without assuming additional matter in the halo.

\section{Dropping the assumption of stationarity} \label{s5}

Unlike GR, NC maintains itself mathematically simple even in absence of Killing vectors or without other simplifying assumptions. We showed in the last Sections how the usual, stationary and axisymmetric case can be solved way more straightforwardly by skipping the full Einstein Equations and employing instead Ehlers' formalism. Here we present, as an example, how even the non-stationary case can be quite easily treated with the same approach.

With the same symmetries employed earlier, but now exploiting $\nabla\times\mathbf{g} = -2\dot{\boldsymbol{\omega}}$, the equation of motion for a $r = $ constant solution constrained on the galactic plane $z = 0$ assumes the following form:
\begin{align} \label{supp v}
    v^2 = a_c r - 2rv\omega_z|_{z = 0}\,, \qquad
    \dot v = a_t\,,
\end{align}
where all quantities can now depend on $t$, in addition to $r$, and $g_{r}|_{z = 0} := -a_c(t,r)$ and $g_{\phi}|_{z = 0} := a_t(t,r)$ are the centrifugal and tangential accelerations, respectively. Note that in the non-stationary case $\mathbf g$ is not the gradient of a scalar potential, so axial symmetry no longer implies that $g_\phi$ vanishes. We have again assumed that there are no particles escape from the equatorial plane,  guaranteed by the planar symmetry (in the sense that was discussed in Section \ref{s3}), and thus $g_z|_{z = 0} = 0$ and $\omega_r|_{z = 0} = 0$. 

The remaining field equations describe a $\boldsymbol{\omega}=\nabla V$ with a Coriolis potential that must be harmonic and is assumed to obey the planar antisymmetry condition $V(t,r,z) = -V(t,r,-z)$. This means that its $z$ component is the only one that does not vanish on the symmetry plane, and we can call it simply $\omega(t,r)=\omega_z|_{z=0}$.

The tangential acceleration $a_t$ is given by the $\hat{z}$ component of $\nabla\times\mathbf{g} = -2\dot{\boldsymbol{\omega}}$ and satisfies the differential equation:
\begin{align}
    (ra_t)_{,r} = -2r \dot\omega\,,
\end{align}
being the other components vanishing on $z=0$. Finally, the modified Poisson equation evaluated at $z = 0$ reads as before:
\begin{align}
    \frac{1}{r}\left(ra_c\right)_{,r} + b = 4\pi G\varrho|_{z = 0} - 2\omega^2\,,
\end{align}
where $g_{z,z}|_{z = 0} := -b(t,r)$.

It can be easily found by quadrature:
\begin{equation} \label{v omega with t}
    (rv(t,r))_{,r} = C(r) - 2r\omega(t,r)\;,
\end{equation}
where $C(r)$ is an arbitrary integration term. This relation between the velocity field and the Coriolis field has no precedents in the stationary case. As a consequence, we cannot now substitute some time-dependent analogues of the $v$'s and $\omega$'s profiles used in Section (\ref{s3c}). Indeed, if:
\begin{equation} \label{Poiss t}
    \omega(t,r) = \omega_0(t)J_0(\gamma(t)r), \quad v(t,r) \sim^{r\rightarrow\infty} v_f(t)\,,
\end{equation}%
are substituted in (\ref{v omega with t}), the asymptotic relation:
\begin{equation}
    C(r) - 2\omega_0(t)r J_0(\gamma(t)r) \sim^{r\rightarrow\infty} v_f(t)\,,
\end{equation}%
can be satisfied only for trivial, time-independent choices of $\omega_0(t), \gamma(t)$ and $v_f(t)$.

This fact can be interpreted by saying that the solution presented in Section \ref{s3c} is not robust for time evolution, although mathematically acceptable. A more robust Newtonian-Cartan system can be obtained with some superposition of the single-harmonic profiles (\ref{single harmonic}).\\
As we mentioned above, for the Coriolis field it can be chosen a superposition analogous to the one built in \cite{Balasin:2006cg}. It may hence decay at infinity as:
\begin{equation}
    \omega(t,r) \sim^{r\rightarrow\infty} -\frac{v_f(t)}{2r}\,,
\end{equation}%
which is consistent with the relation (\ref{v omega with t}), with a $C(r)$ tending to zero at infinity. This Coriolis field can support flat rotation curves, according to (\ref{supp v}), even if $a_c$ decays faster than $1/r$. Finally, $b(t,r)$ can be obtained from the last equation (\ref{Poiss t}), analogously to what we did in Section \ref{s3c}.

\section{Discussion and conclusion} \label{concl}

In this paper, we address Ehlers' FT in the limit $\lambda \to 0$. This does not recover Newtonian gravity, but rather Newton-Cartan gravity. A prominent feature of this theory is a second vector gravitational field $\boldsymbol{\omega}$ named the Coriolis field. We show how this new degree of freedom is related to that part of the metric field that generates frame-dragging effects.

We apply NC for axially symmetric configurations of matter, having in mind models for disc galaxies, and studied the possibility that flattish rotation curves could be naturally predicted in this context. We also show that some models in the literature that effectively describe the dynamics of galaxies, and of their rotation curves, such as the $(\eta,H)$ model, can be framed in the limit $c \to \infty$ in the NC theory with a non-vanishing Coriolis field, which is tightly related to the quasi-local angular momentum \L$_D$ of background spacetime.

NC gravity theory has the great advantage of being mathematically much simpler than GR. Therefore, some of the standard assumptions adopted in the latter theory for describing some physical systems can be discarded. One prominent example is the stationarity assumption. Calculations without it are almost impossible to carry out in GR, instead being rather simple in NC theory as we show in Section \ref{s5}. Analogously, in future work the axisymmetry assumption may be abandoned, to develop an even more realistic galactic model with Coriolis field. We may expect such non-stationary, non-axisymmetric models to be a perturbative development of what we have shown in Section \ref{s4}, with the perturbations of the form of pressure waves, thus describing the galactic spirals.

Our investigation confirms that the PN formalism might not grasp the full geometric content of GR, failing to properly address the frame-dragging effects that account for the existence of the Coriolis field $\boldsymbol{\omega}$ even in the limit $c\to \infty$, where they are expected, in the standard approach, to be negligible. This is highlighted for the cases where the background geometry is not Minkowskian.

The drawback of NC is the same as that of Newtonian theory. The limit $c\to\infty$ amounts to treating the physical system at hand as virtually of vanishing dimension and with infinite time scale, which might not be appropriate for analyzing systems such as galaxies. The effects of retarded potentials, for example, \cite{Carati:2008bmr, Re:2020oiq, Glass:2024zgy}, also cannot be appreciated in this limit. We note that, since in NC theory the gravitational potentials travel at infinite speed, this theory is not reliable for a perturbative study, such as the one mentioned above regarding galactic spirals, since the perturbations of the potentials propagate too quickly compared to the size of a galaxy.

The loss of causality in NC gravity is discussed, e.g., in Section 3 of \cite{BuchertFT}, and can be particularly appreciated in the Ehlers' example of the NC limit of a gravitational wave metric, presented in both \cite{Ehlers:2019aco} and \cite{ehlers1997examples}. In this sense, our work might be considered complementary to \cite{Galoppo:2024vzo}, where the authors propose a quantitative way to evaluate the reliability of the PN expansion.

In future perspective, it will be interesting to analyze in some more detail NC applied to the galactic context. In particular, to find physical solutions for the Coriolis field that merge for different galaxies, we should consider multi-galaxies models. The study of systems with no simplifying symmetries is made easier by employing NC. Moreover, to overcome the drawback mentioned earlier, it would be interesting to perform a $\lambda$ expansion of Ehlers' FT. This could perhaps allow us to create a bridge with \cite{Galoppo:2024vzo} and perhaps find a relation $\tilde\alpha \propto \lambda^2$.

Even in first order in $\lambda$, a source term arises for the curl of the Coriolis field; it is indeed the source in the usual, linearized gravitomagnetic equations, and it is proportional to the angular momentum density \cite{Ludwig:2021kea, AstesianoRuggiero1, AstesianoRuggiero2, Srivastava2023, LeCorre:2024tga}. Moreover, in the exact non-linear gravitomagnetic equations \cite{LandauLifshitz, NLGM1, Natario2007, costa_gravito-electromagnetic_2014, NLGM3, Costa1, Costa:2023awm} even other non-linear terms arise to source the curl of $\boldsymbol{\omega}$ and the divergences of $\boldsymbol{\omega}$ and $\textbf{g}$. Since for small but non-vanishing $\lambda$, the Coriolis field equations are no more sourceless, neither linear, the discussion about the boundary conditions can be substantially altered and rendered more physical. The $\lambda \neq 0$ generalization also overcomes the problem of the instantaneity of the theory, and phenomena related to the presence of retarded gravitational potentials can be addressed.

\medskip
\textbf{Acknowledgments}\\
The authors thank Sergio L. Cacciatori and Vittorio Gorini for useful discussions and comments. They are also grateful to Marco Bruni and Marco Galoppo for calling our attention to Ehlers' work \cite{ehlers1997examples}.

\bibliographystyle{dinat}
\bibliography{Ehlers.bib}

\end{document}